%% file: paper.tex
\documentclass[letter,prd,aps,twocolumn,floatfix,superscriptaddress]{revtex4}

\usepackage{amssymb,amsmath,graphicx}
\usepackage{hyperref}
\usepackage[usenames,dvipsnames]{color}
\usepackage{array}
\makeatletter
\renewcommand*{\p@section}{\S\,}
\renewcommand*{\p@subsection}{\S\,}

\makeatother

\input{aas_macros.tex}

\begin{document}

\date{\today}
\title{Black Hole Dynamics in Einstein-Maxwell-Dilaton Theory}

\author{Eric W.~Hirschmann}
\affiliation{Brigham Young University, Provo, UT 84602, USA}
\author{Luis Lehner}
\affiliation{Perimeter Institute for Theoretical Physics, Waterloo, ON, N2L2Y5, Canada}
\author{Steven L. Liebling}
\affiliation{Long Island University, Brookville, New York 11548, USA}
\author{Carlos Palenzuela}
\affiliation{Universitat de Iles Baleares, Spain}


\begin{abstract}
We consider the properties and dynamics of black holes within a family of
alternative theories of gravity, namely Einstein-Maxwell-dilaton~(EMD)
theory.  We analyze the dynamical evolution of
individual black holes as well as the merger of binary black hole systems.
We do this for a wide range of parameter values for the family of EMD
theories, investigating, in the process, the stability of these black holes.
We examine radiative 
degrees of freedom, explore the impact of the
scalar field on the dynamics of merger and compare with other scalar-tensor
theories.  We argue that the dilaton can largely be discounted in understanding
merging binary systems and that the endstates essentially interpolate between 
charged and uncharged, rotating black holes. 
For the relatively small charge 
values considered here, we conclude that these black hole systems will be 
difficult to distinguish from their analogs within general relativity.  
\end{abstract}

\maketitle


\section{introduction}
One particularly exciting prospect arising from
the recent advent of gravitational wave astronomy
is the possibility of testing General Relativity~(GR) in the previously inaccessible
strongly gravitating/highly dynamical regime. Indeed, first steps in this direction have
already been enabled by the three available detections (so far), 
GW150914~\cite{Abbott:2016blz}, GW151226~\cite{Abbott:2016nmj}, and GW170104~\cite{Abbott:2017vtc}. Analysis
of these signals reveals that they are consistent with those produced by black hole 
mergers in GR~\cite{TheLIGOScientific:2016src,Yunes:2016jcc}, with independent
and complementary tests coming from the inspiral and post-merger periods. 

Ongoing efforts with additional detections and studies of predicted signals will
allow for further scrutiny~\cite{Meidam:2014jpa,Berti:2015itd,Yang:2017zxs}. 
Accurate predictions of possible signals are not only important
to aid in future detections but are also helpful in realizing important tests of the theory.
Such tests could potentially indicate that nature deviates from GR. However,
simply identifying possible deviations is unlikely to provide sufficient 
guidance as to the alternative that nature may have chosen. In contrast to 
the GR case, our understanding of potential signals within alternative 
theories of gravity is still rather limited. 

To date, this approach of looking for such deviations has been primarily
restricted to the consideration of phenomenological models
(e.g.~\cite{Yunes:2009ke,Agathos:2013upa}).  Another avenue for obtaining
detailed predictions is to use fully nonlinear treatments within specific
alternative gravitational theories. Such an approach, however, requires
theories which possess well posed evolutionary problems in addition to
producing spacetime deviations in, for example, binary black hole mergers.
Such a requirement is fairly stringent and limits the number of possible
options.  We do note that there is a body of previous work which has studied
the possibility of deviations within the context of binary neutron systems (see
e.g.~\cite{Barausse:2012da,Shibata:2013pra,Berti:2013gfa,Palenzuela:2013hsa}).
However, currently available observations indicate that the majority of
events that we might expect in the near future should correspond to binary
black hole systems~\cite{Abbott:2016nhf}. Recent approaches to
the nonlinear regime
for such binary mergers in GR alternatives are beginning to address concerns
with, for example, ill-posedness~\cite{Okounkova:2017yby,Endlich:2017tqa,Cayuso:2017iqc}.

In the present work we study black hole systems (single and binaries) in the 
Einstein-Maxwell-dilaton~(EMD) theory~\cite{Garfinkle:1990qj}. This theory, 
originating from a low energy limit of string theory, allows for black holes 
that have mass, rotation, charge and scalar ``hair'' together with
scalar, vector and tensor radiative channels. Furthermore, its mathematical 
structure allows for defining a well posed initial value problem. It therefore 
offers an interesting theoretical and computational playground within which 
to explore possible deviations from the standard model (i.e.  GR) prediction. 

While we provide below some specifics regarding the black holes we consider
in this work, it is worth mentioning that our understanding of black hole
systems in EMD is admittedly rather limited.  For example, analytic solutions
are known primarily for non-rotating systems.  
A spherically symmetric family 
of solutions exists parametrized by the scalar (dilaton) coupling, 
and these solutions are known analytically across a range of coupling values.  However,
analytic investigations of
their perturbations, stability, and rotating generalizations are at best
limited to a handful of specific values of the coupling.

While our aim in the current work is not necessarily to address all of these
questions, we do study single and binary black hole systems within EMD and
draw some general conclusions to help understand
the dynamics of coalescing binaries.

The subsequent presentation is divided up as follows. Section~II presents the equations of motion describing the systems' dynamics.
Section~III includes a brief description of known, non-spinning black holes, possible instabilities, and a discussion
of possible radiative effects. Section~IV presents results for both single and binary black hole systems for the
case of small charge. We conclude in Section~V. We defer to an appendix a description of EMD black hole solutions
in isotropic coordinates and to a second appendix a calculation of the radiative properties
in the Jordan frame.

\section{Equations} 

The alternative theory of gravity that we consider has origins in low energy 
string theory.  A particular sector of this theory includes a U(1)
gauge field and a scalar field that couples exponentially to the gauge field.
For definiteness, we consider the action for low energy, heterotic string 
theory
\begin{equation}
S = \int d^4 x \sqrt{-{\tilde g}}\, e^{-2\phi} \left[ R +\! \Lambda +\! 4 \bigl( \nabla \phi \bigr)^2\! -\! F^2 -\! {H^2\over12} \right]  
\end{equation}
where the matter content includes a scalar field serving as the dilaton, $\phi$, a U(1) gauge field, $F_{ab}$, and a three-form field, $H_{abc}$, which is 
related to the axion and which, together with the cosmological constant $\Lambda$, 
we set to zero in the following.  This action is written
with respect to the string metric, ${\tilde g}_{ab}$, which is the metric
to which strings couple.  (It is also referred to as the Jordan metric or 
frame.) In many treatments, including this work, a 
conformal transformation is performed to the Einstein metric, or frame, 
via $g_{ab} = e^{-2\phi} {\tilde g}_{ab}$.  
On performing this transformation at the level of the action, we arrive
at the expression
\begin{equation}
S = \int d^4 x \sqrt{-g} \left[ R - 2 \bigl( \nabla \phi \bigr)^2 - 2 V - e^{-2\alpha_0\phi} F^2 \right]  
\end{equation}
where, as before, the scalar field, $\phi$, is the dilaton and $F_{ab}$ is the 
Maxwell tensor.  Note that we have chosen to generalize the theory a bit by
including $V(\phi)$, a potential for the dilaton, together with including 
the real constant $\alpha_0$ to parameterize among theories.  In particular, 
$\alpha_0=0$ is just Einstein-Maxwell minimally coupled to a real 
scalar field, $\alpha_0=1$ is the sector of low energy string theory referred
to above, and 
$\alpha_0=\sqrt{3}$ corresponds to Kaluza-Klein theory~\cite{Frolov:1987rj,Horne:1992zy}.  This action defines a class of theories often referred to as Einstein-Maxwell-dilaton.  Our interest focuses on  
dynamical processes within this theory and how they might compare with standard 
general relativity.  

The equations of motion that follow from this action are the Einstein-Maxwell 
equations coupled nonlinearly to a propagating scalar field (the dilaton),
namely   
\begin{eqnarray} 
R_{ab} & = &  2 \left( T_{ab} -\frac{1}{2} g_{ab} T \right) \\ 
\nabla^a \nabla_a \phi & = & {1\over2} {\partial V \over \partial \phi} - {\alpha_0 \over2} \, e^{-2\alpha_0\phi} \, F^2 \label{eq:dilaton}\\ 
\nabla^a F_{ab} & = &  -I_b ~~.
\end{eqnarray} 
Notice that the exponential coupling of the dilaton in its equation of motion
is again present in both the four-current, $I_b$, and the stress-energy tensor 
\begin{eqnarray} 
  I_b &=& - 2\alpha_0 \nabla^a \phi F_{ab} \\
  T_{ab} &=& T^{\phi}_{ab} + e^{-2 \alpha_0 \phi} T^\mathrm{EM}_{ab} \\
  T^{\phi}_{ab}  &=& \nabla_a \phi \nabla_b \phi 
  - \frac{1}{2} g_{ab} \left[ \nabla_c \phi \nabla^c \phi + V(\phi) \right] \\
    T^\mathrm{EM}_{ab}  &=& F_{ac} {F_b}^c - \frac{1}{4} g_{ab} F^2         .
\end{eqnarray}

These equations are supplemented by the identity $\nabla_{[a} F_{bc]} = 0$.  
Because of the presence of this and related constraint equations in the above
evolution system, we, in fact, 
use an extended Maxwell system which aids in damping dynamically these constraints.  To the 
above Maxwell equations, we add extra scalar fields, $\Psi$ and $\Phi$, in 
such a way that the 
Maxwell constraints are allowed to propagate at the speed of light.  These 
equations become 
\begin{eqnarray} 
\label{eq:maxwell}
\nabla^a \bigl( F_{ab} + g_{ab} \Psi \bigr) & = & \kappa_1 n_b \Psi 
- I_{b} \\  
\nabla^a \bigl( (*F)_{ab} + g_{ab} \Phi \bigr) & = & \kappa_2 n_b \Phi
\end{eqnarray} 
where the $\kappa$s are real constants used to adjust the constraint damping 
and $(*F)_{ab} = {1\over2} \epsilon_{abcd} F^{cd}$ is the dual of $F_{ab}$.  

We use the usual Cauchy, or 3+1, decomposition in which the spacetime is 
foliated into spacelike hypersurfaces, $\Sigma_t$, labeled by a coordinate 
time, $t$.  The timelike normal is $n^a$ with orientation such that 
$n_a = -\alpha \, \delta_a^t$ and the metric on the hypersurfaces is 
$\gamma_{ij}$.  The lapse and vector shift of the coordinates are given by 
$\alpha$ and $\beta^i$.  The line element of the spacetime is then 
\begin{equation} 
{\rm d}s^2 = - \alpha^2 {\rm d}t^2 + \gamma_{ij} \bigl( {\rm d}x^i + \beta^i {\rm d}t \bigr) \bigl( {\rm d}x^j + \beta^j {\rm d}t \bigr) .
\end{equation}
Note that we define a derivative operator, $D_i$, built
from the 3-metric $\gamma_{ij}$ which should be compared with the full 
derivative operator, $\nabla_a$, built from $g_{ab}$.  Likewise, we define 
a 3-Levi-Civita antisymmetric tensor density as $\epsilon_{bcd} = n^a \epsilon_{abcd}$.  

With respect to these 3+1 variables, we can write the above matter equations 
as  
\begin{eqnarray}
\bigl(\partial_t - {\cal L}_\beta \bigr) \phi & = & - \alpha \Pi 
\label{eom:first}\\
\bigl(\partial_t - {\cal L}_\beta \bigr) \Pi & = & - D^i\bigl( \alpha D_i \phi \bigr) + \alpha K \Pi + {\alpha\over2} \, {\partial V \over \partial \phi} 
  \cr 
  & & \quad - \alpha_0 \, \alpha e^{-2\alpha_0\phi} \bigl[ B_i B^i - E_i E^i \bigr] \\ 
\bigl( \partial_t - {\cal L}_\beta \bigr) E^i \! & = & \epsilon^{ijk} D_j \bigl( \alpha B_k \bigr) + \alpha \bigl[ K E^i - D^i \Psi \bigr] 
  \cr 
  & & \quad - 2 \alpha_0 \, \alpha \bigl[ \epsilon^{ijk} D_j \phi B_k + \Pi E^i \bigr] \\ 
\bigl( \partial_t - {\cal L}_\beta \bigr) \Psi & = & \alpha \bigl[ 2\alpha_0 D_j \phi E^j - D_j E^j - \kappa_1 \Psi \bigr] \\ 
\bigl( \partial_t - {\cal L}_\beta \bigr) B^i \! & = & \! - \epsilon^{ijk} D_j \bigl( \alpha E_k \bigr) + \alpha \bigl[ K B^i + D^i \Phi \bigr] \\ 
\bigl( \partial_t - {\cal L}_\beta \bigr) \Phi & = & \alpha \bigl[ D_j B^j - \kappa_2 \Phi \bigr]  
\label{eom:last}
\end{eqnarray} 
where we define the electric, $E_i \equiv \gamma_i{}^j F_{jc} n^c$, and magnetic, 
$B_i \equiv {1\over2} \epsilon_{ijk} F^{jk}$, fields relative to their projections 
into the spacelike hypersurface, $\Sigma_t$.  Note, too, that we have defined  
$\Pi \equiv - n^a \nabla_a \phi$ and invoked the Lie derivative along the 
shift, ${\cal L}_\beta$.   

For the evolution of the gravitational field in the Einstein frame, we use
the BSSN formalism for which we give here only the projections of the matter 
stress tensor, namely 
\begin{eqnarray} 
\rho = n^a n^b T_{ab} &=& D_i\phi D^i \phi + \Pi^2 + V  
\label{eom:sources1}
\nonumber \\
&+& e^{-2\alpha_0\phi} \bigl( B_{i} B^{i} + E_i E^i \bigr)   \\
j_i  = - n^a \gamma_i{}^b T_{ab} &=&   - 2 \Pi D_i\phi - 2 e^{-2\alpha_0\phi} \epsilon_{ijk} E^j B^k \\  
S_{ij} = \gamma_i{}^a \gamma_j{}^b T_{ab} &= & 2 D_i \phi \, D_j \phi + 2 e^{-2\alpha_0\phi} \, \bigl( B_i B_j  \, - E_i E_j  \bigr) \nonumber \\
  & - &  \gamma_{ij} \, \Bigl[ D^k\phi \, D_k\phi - \Pi^2 + V  
\nonumber \\
  & & \qquad + e^{-2\alpha_0\phi} \, \bigl( B_{k} B^{k} - E_k E^k \bigr) \Bigr].
\label{eom:sources2}
\end{eqnarray} 

\begin{figure}[h]
\centering
\includegraphics[width=8.2cm,angle=0]{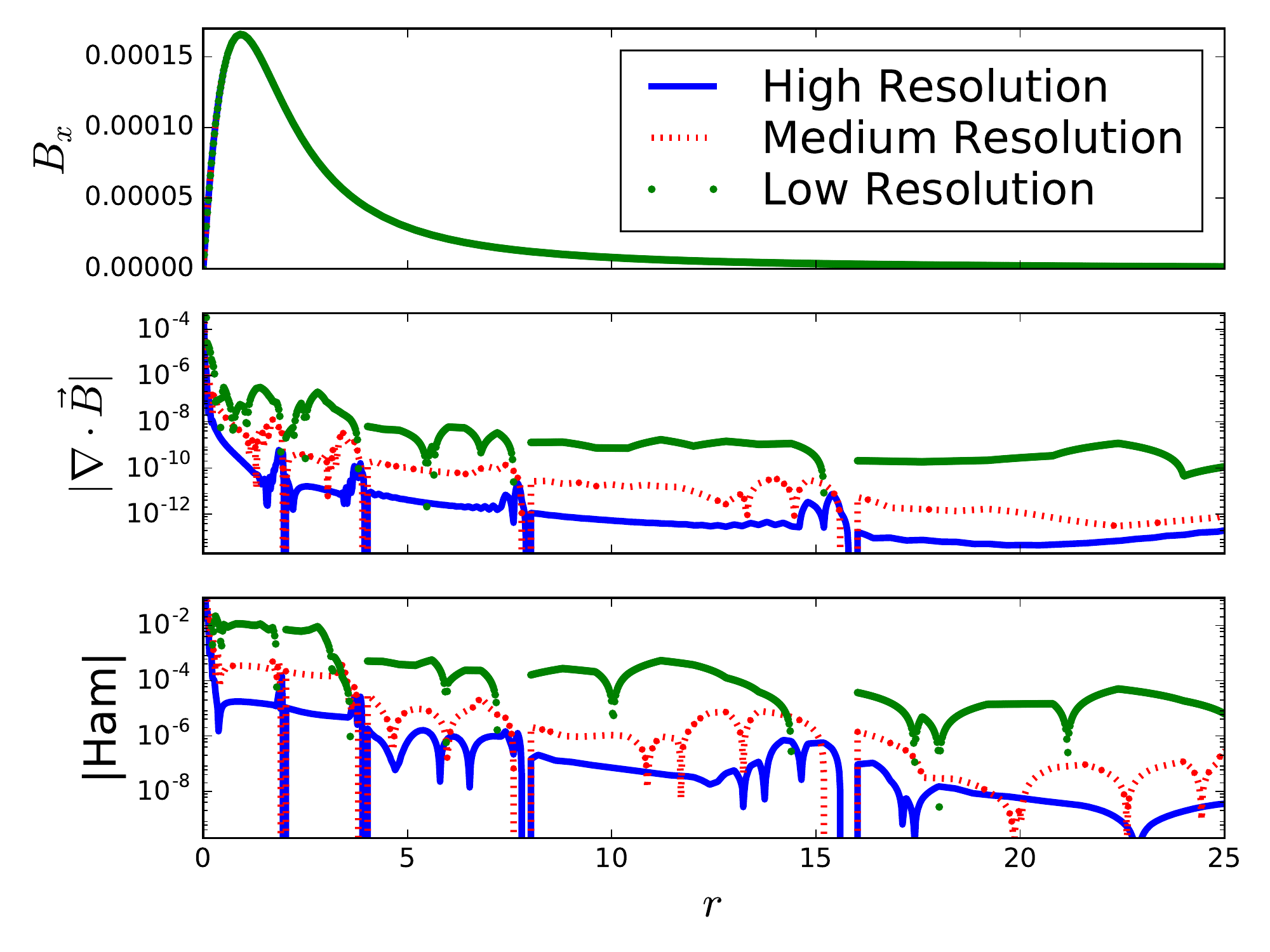}
\caption{ Convergence for a magnetic black hole. Shown are various fields at late time ($t=80M$) for three different resolutions. Each run uses fixed mesh refinement~(FMR), but differs by a factor of 2x in resolution for the same grid structure. 
The top panel shows that all three resolutions approach the same, static solution. The bottom two panels are measures of errors demonstrating that higher resolution runs have lower errors. In particular, the middle frame shows the
divergence of the magnetic field which should be zero except at the center (because of the monopole charge).
The bottom frame shows the residual of the Hamiltonian constraint.
} 
\label{fig:convergence}
\end{figure}

We implement the resulting equations using techniques described 
previously~\citep{Anderson:2006ay,Palenzuela:2006wp,Liebling,Anderson:2007kz,B_fields_are_us,had_webpage}, and demonstrate convergence for a particular 
case as seen in Fig.~\ref{fig:convergence}. We note that we adopt 
the ``1+log'' and Gamma drivers for the lapse function and the shift vector.
Finally, using standard BSSN notation, we set the quantities $K_0$ and $\eta$ 
to $0$ and $2/M \; (1/M)$, respectively, 
for the single (binary) black holes studied here~\cite{Alcubierre:2002kk}, 

\section{Preliminaries}\label{preliminaries}
Before discussing the dynamics as revealed by numerical evolutions,
we first discuss some properties of the model.

\subsection{Known black hole solutions}\label{knownsolutions}

There exist a number of black hole solutions in EMD.  One such solution is a 
static, magnetically charged black hole solution found 
in~\citep{Gibbons:1987ps,Garfinkle:1990qj}.  In Schwarzschild-like 
coordinates, this solution takes the form 
\begin{eqnarray}\label{eqnssch}
{\rm d}s^2 & = & - \Bigl( 1 - {r_{+} \over r} \Bigr) \Bigl( 1 - {r_{-} \over r} \Bigr)^{1-\alpha_1} \, {\rm d}t^2 
  \cr 
  & & + \Bigl( 1 - {r_{+} \over r} \Bigr)^{-1} \Bigl( 1 - {r_{-} \over r} \Bigr)^{\alpha_1-1} \, {\rm d}r^2 
  \cr 
  & & + r^2 \Bigl( 1 - {r_{-} \over r} \Bigr)^{\alpha_1} {\rm d}\Omega^2 \\
    F_{\theta\phi} & = & Q_m \sin\theta \\ 
    e^{-2\alpha_0\phi} & = & e^{-2\alpha_0\phi_0} \, \Bigl( 1 - {r_{-} \over r} \Bigr)^{\alpha_1} 
\end{eqnarray}
where $\alpha_1 = 2 \alpha_0^2 / (1+\alpha_0^2)$ and $\phi_0$ is the 
asymptotic value of the dilaton at spatial infinity.  This solution 
has magnetic charge $Q_m$.  The constants $r_{\pm}$ are given 
in terms of $Q_m$, $\phi_0$, and the ADM mass, $M$, of the spacetime:
\begin{eqnarray}
2M & = & r_{+} + \bigl( 1 - \alpha_1 \bigr) r_{-} \\ 
2 Q_m^2 & = & e^{2\alpha_0 \phi_0} \, r_{+} r_{-} \bigl( 2 - \alpha_1 \bigr) .
\end{eqnarray}
Properties of this black hole solution are given, for instance, 
in~\citep{Garfinkle:1990qj,Horowitz:1992jp}.  For our purposes, it suffices 
to note that $r_{+}$ corresponds to an event horizon and $r_{-}$ is the 
location of a 
curvature singularity.  Note, too, that the dilaton, for $r>r_{-}$, is 
larger than its asymptotic value and monotonically decreases towards $\phi_0$.

There is a discrete electromagnetic duality in this theory that exchanges
magnetic and electric solutions.  The explicit duality leaves the metric 
unchanged, but sends $F_{ab} \rightarrow e^{-2\alpha_0\phi} (*F)_{ab}$ and 
$\phi \rightarrow -\phi$.  Because of the presence of the dilaton, 
the electrically charged solution is, in fact, a different solution.  While
the metric takes the same form as above, the Maxwell field and dilaton become 
\begin{eqnarray}
F_{tr} & = & { Q_e \over r^2} \\ 
e^{2\alpha_0\phi} & = & e^{2\alpha_0\phi_0} \, \Bigl( 1 - {r_{-} \over r} \Bigr)^{\alpha_1} ,
\end{eqnarray}
and for which the constants $r_{\pm}$ satisfy 
\begin{eqnarray}
2M & = & r_{+} + \bigl( 1 - \alpha_1 \bigr) r_{-} \\ 
2 Q_e^2 & = & e^{-2\alpha_0 \phi_0} \, r_{+} r_{-} \bigl( 2 - \alpha_1 \bigr) .
\end{eqnarray}
In this case, the solution has electric charge $Q_e$, ADM mass $M$, asymptotic
dilaton value, $\phi_0$, an event horizon at $r=r_{+}$, a curvature
singularity at $r_{-}$, and, for $r>r_{-}$, a dilaton monotonically increasing 
towards $\phi_0$.  

We know of
linearized perturbations of these black holes only for the case 
$\alpha_0=1$~\cite{Ferrari:2000ep}. There, the quasi-normal mode
spectra has been computed, and it was shown that the presence of the dilaton induces
a difference in the spectra of axial and polar perturbations. This difference 
breaks
isospectrality, a property known to apply to  both Schwarzschild and Reissner-Nordstr{\"o}m black
holes.

Note that in the limit $\alpha_0 \rightarrow 0$ the solution corresponds to
that of a charged RN black hole.  As $\alpha_0 \rightarrow \infty$,
the solution is simply an uncharged Schwarzschild black hole for which the
Maxwell field is zero and the dilaton is a constant.  By extension we
expect (and show below)
that for the rotating solutions $\alpha_0$ 
interpolates between the charged Kerr-Newman black hole and the uncharged Kerr solution,
the latter unadorned by scalar or vector fields.
One way of understanding the $\alpha_0\rightarrow\infty$ limit is that the
gravitational sector decouples from the matter sector (see, for example,~\cite{Bizon:1992gi}). In this limit, regardless
of the behavior of the dilaton and Maxwell field, the gravitational solutions
are just those of GR, such as Schwarzschild and Kerr black holes.

Moving away from spherical symmetry, there are rotating black hole solutions
in EMD, but to our knowledge, an analytic solution is only known for 
the case $\alpha_0 = \sqrt{3}$ corresponding to 
Kaluza-Klein~\cite{Frolov:1987rj}.  Rotating solutions in EMD for other 
coupling values have been constructed 
numerically~\cite{Kleihaus:2002tc,Kleihaus:2003sh,Kleihaus:2003df}.  
Further, the behavior of perturbations 
and questions related to stability would appear to be largely unexplored.  An
important exception to this is~\cite{Ferrari:2000ep} which considers the 
quasi-normal modes of the spherically symmetric solutions for $\alpha_0=1$. 
Recently, time dependent, spherically symmetric solutions sourced by a charged 
null dust flow have been presented in EMD~\cite{Aniceto:2015klq}. 
Interestingly, it is only for the coupling $\alpha_0=1$ that the  
solution describes a time-dependent dilaton (in addition
to time-dependent metric and gauge fields).

In what follows we study both single and binary black hole scenarios and
derive general statements about the dynamical behavior induced by EMD. 
As a prelude, we first present a simple-minded picture which can capture
a possible transition in the behavior of the scalar field.

\subsection{Scalar field instabilities}\label{instab}

Here we argue two types of instabilities could trigger non-trivial
behavior of the scalar field. The analysis follows closely the
arguments presented in~\cite{Cardoso:2013fwa}. 

Consider the linearized equation of motion for the dilaton,  Eq.~(\ref{eq:dilaton}), which becomes
\begin{equation}
\Box \phi = -\frac{\alpha_0}{2} (1 - 2 \alpha_0 \phi) F^2
\end{equation}
with $\Box$ defined with respect to a background metric. We can rewrite
the above equation as 
\begin{equation}
(\Box - \mu^2) \phi = -\frac{\alpha_0}{2} F^2 
\end{equation}
by introducing
$\mu^2 \equiv \alpha_0^2 F^2$. Following the discussion
in Appendix~\ref{knownsolutions}, notice that $\mu^2$ can have
either sign, depending on whether the magnetic or electric field dominates.
Assuming the charge $Q$ (either magnetic or electric case) is small, the metric is
that of Schwarzschild to ${\cal O}(Q^2)$. 
Expanding the dilaton in spherical harmonics as
 $\phi = \Sigma_{lm} e^{-i\omega t} Y_{lm}(\theta,\varphi) \Phi_{lm}(r)/r$
yields
\begin{equation}
f^2 \Phi_{lm}^{''} + f' f \Phi_{lm}^{'} + \left[\omega^2 - f {\bar V}(r)\right] \Phi_{lm} = -f\, \frac{\alpha_0}{2} \,F^2\, r e^{i \omega t}
\end{equation}
with
${\bar V}(r) = l(l+1)/r^2 + 2M/r^3 + \mu^2$ and $f(r) = 1 - 2M/r$. This equation is similar to that
obtained 
in~\cite{Cardoso:2013fwa} except for the presence of a source term on the right hand side
which is independent of $\Phi_{lm}$.
Integrating the equation over a period does away with the source,
 and in this cycle-averaged sense
we will ignore it in the following.
A sufficient condition for an instability is that $\int_{r_\mathrm{BH}}^{\infty} {\bar V}(r) dr < 0$ which translates into
\begin{equation}
\int_{r_\mathrm{BH}}^{\infty} \left[\frac{l (l+1)}{r^2}  + \frac{2M}{r^3} \right] dr < -\alpha_0^2 \int_{r_\mathrm{BH}}^{\infty} F^2 dr ,
\end{equation}
and thus the instability condition becomes
\begin{equation}
\frac{2 l (l+1) + 1}{4 M} < -\alpha_0^2 \int_{r_\mathrm{BH}}^{\infty} F^2 dr.  \label{spontscalcond}
\end{equation}

Condition Eq.~(\ref{spontscalcond}) implies that the electrically dominated case
is subject to this instability while the magnetically dominated case is not.
Indeed, such an instability resembles the standard negative mass instability, 
and that will be our primary focus here.  However, we note that the 
magnetically 
dominated case may be subject to a superradiant type instability associated
with rotating black holes as the effective mass $\mu$ could introduce a 
potential barrier that would provide feedback for such a process. 

For now, consider equation (\ref{spontscalcond}) and evaluate it in a simple
case such as the solution provided in Appendix~\ref{knownsolutions}  
concentrating 
on the electrically dominated case and for small charge. Evaluation of 
Eq.~(\ref{spontscalcond}) gives
\begin{equation}
\frac{2 l (l+1) + 1}{4 M} < \alpha_0^2 \frac{Q_e^2}{12 M^3}.
\end{equation}
for which the smallest bound on $\alpha_0$ is achieved with $l=0$. Clearly 
the coupling and charge must satisfy 
\begin{equation}
3 < \alpha_0^2 \left( \frac{Q_e}{M} \right)^2.
\end{equation}
On rearranging, this condition, $\alpha_0 > \sqrt{3} (M/Q_e)$, indicates
an instability at a large value of $\alpha_0$ for
small charges. This analysis suggests that one needs 
``large parameters to get large effects." 
For concreteness, a charge of $Q_e/M =10^{-3}$ predicts an instability for 
$\alpha_0 \gtrsim 1.7 \times 10^{3}$. 

\subsection{Extra degrees of freedom and black hole binaries}

Although EMD is interesting in its own right, the remarkable
direct detections of the mergers of black hole binaries by LIGO
provides arguably the first opportunity to truly test gravity in strong
field/highly dynamical settings. As such, we can regard EMD as an alternative 
to GR, one that includes additional degrees of freedom in the form of a scalar 
and vector field.
In that respect, EMD has a scalar degree of freedom in common with scalar tensor
theories where several new phenomena have been well established. 

In such theories one can characterize the scalar field by its
\textit{scalar monopole charge}. This scalar charge (one generally drops
the ``monopole'') 
can be evaluated by computing the divergence of the field over some large Gaussian surface.
In particular, at large
radius one considers the behavior of the scalar as $\phi(r) \approx \phi_0
+ \phi_1/r$, so that $\phi_1$ is the scalar charge and $\phi_0$ the asymptotic value of 
the scalar field.

In some scalar tensor gravity theories it has been found that the scalar field 
can grow significantly around compact neutron stars~\cite{PhysRevLett.70.2220,Damour:1996ke}. This process, known 
as {\em spontaneous scalarization}, induces a scalar charge around each NS 
that determines the extent to which the
theory's predictions differ from those of GR~\cite{PhysRevLett.70.2220,Damour:2010rp}. 
In particular, such effects, originating in the scalar field, allow for an 
enhancement of the gravitational force and for
additional channels of radiation (such as dipole
scalar radiation).  As a result, one generally expects such binaries to merge earlier 
than their GR counterparts (e.g.~\cite{Damour:1996ke,Barausse:2012da,Palenzuela:2013hsa,2014grav.book.....P}). 
Black holes in such
theories are identical to those of GR and the gravitational 
waves observed in their merger provide no new features and therefore offer no distinguishing test of the theory (unless the asymptotic value of the scalar field is time-dependent~\cite{Horbatsch:2011ye}).

In contrast, the EMD gravity theory that we study here
allows for a scalar charge even without the presence of matter as long as the gauge field (and thus
the gauge charge of the black hole) is non-zero. 
There are two ways to interpret the U(1) gauge field.
Astrophysically, one expects the black hole
gauge charge to be very small if the gauge field corresponds to the Standard Model (SM) electromagnetic field.
On the contrary, one can consider this gauge field not as the usual electromagnetic field of the SM, but instead an additional field that is simply a component
of gravity.
In that case, 
there exist no constraints in principle
for the black hole charge, but it is natural to expect it should also be small.
Regardless of these considerations, one can consider black holes in EMD theory as natural
proxies to consider general fields describing gravity; the spin 0
scalar, the spin 1 gauge field, and the spin 2 metric field.

\section{Results}
In what follows we discuss results obtained for single and binary
black hole systems, which have been studied via numerical simulations. Our implementation
of Eqs.~(\ref{eom:first})-(\ref{eom:last}), along with the BSSN equations
coupled to the sources in Eqs.~(\ref{eom:sources1})-(\ref{eom:sources2}),  adopt finite difference techniques satisfying summation
by parts on a regular Cartesian grid~\cite{SBP2,SBP3}. All fields are discretized
using a fourth order accurate scheme. The time evolution of the resulting equations 
is performed by using a third order accurate Runge-Kutta scheme~\cite{Anderson:2006ay,Anderson:2007kz}. 
We employ adaptive mesh refinement~(AMR) via the HAD computational infrastructure. This
provides distributed, Berger-Oliger style AMR~\cite{had_webpage,Liebling} with
full sub-cycling in time, with the inclusion of an improved treatment of artificial
boundaries~\cite{Lehner:2005vc}. In addition, for cases with large values of $\alpha_0$, and for which 
the field becomes fairly non-smooth in the central area
of each black hole, we adopt a more aggressive form of dissipation
(essentially a high-pass filter) on the dilaton localized to those
regions which lie well within the apparent horizon. 

In our single black hole runs we adopt a mass $M=1$ and employ a Cartesian grid
with extent $[-64,64]^3$ with a base resolution of $81$ equispaced points
in each direction. There are $6$ levels of concentric, fixed, finer meshes covering half the extent of each parent, such that
the finest resolution is $h=0.05$.
For these runs, we employ the standard gauge
driver with $\eta=2/M$ where $\eta$ is the standard gauge parameter of the
Gamma driver coordinate choice.

In our binary black hole runs, we adopt $m_1=m_2=1/2$ for the equal mass
case or for the unequal case $m_1=0.5788,m_2=0.3852$. Our coarsest computational grid is defined
over $[-204.8,204.8]^3$ and each direction is covered with $81$ uniformly spaced points.
We additionally employ eight levels of refinement (with a $2:1$ refinement
ratio). The first two are fixed in $[-102.4,102.4]^3$ and $[-34,34]^3$ while the remaining 
six adapt dynamically through the shadow hierarchy, giving a finest resolution of
$h=0.02$. For these runs, we use the standard gauge
driver but find that resolving the rapid dynamics of the merger required choosing $\eta=1/(m_1+m_2)$.

\subsection{Single black holes}

We present first the 
behavior of the dilaton scalar field in spacetimes with a weakly charged single black hole.
As a simple way to choose initial data and to explore the stability of the black hole,
we choose for our initial data a black hole solution in GR to which we add a monopole electric
(or magnetic) field (whose asymptotic charge value is kept fixed). The dilaton is set to a constant equal to 
its asymptotic value, $\phi_0$.  This data does not correspond to a stationary solution
but it is consistent with the constraints to ${\cal O}(Q^2)$.

\begin{figure}[h]
\centering
\includegraphics[width=8.0cm,angle=0]{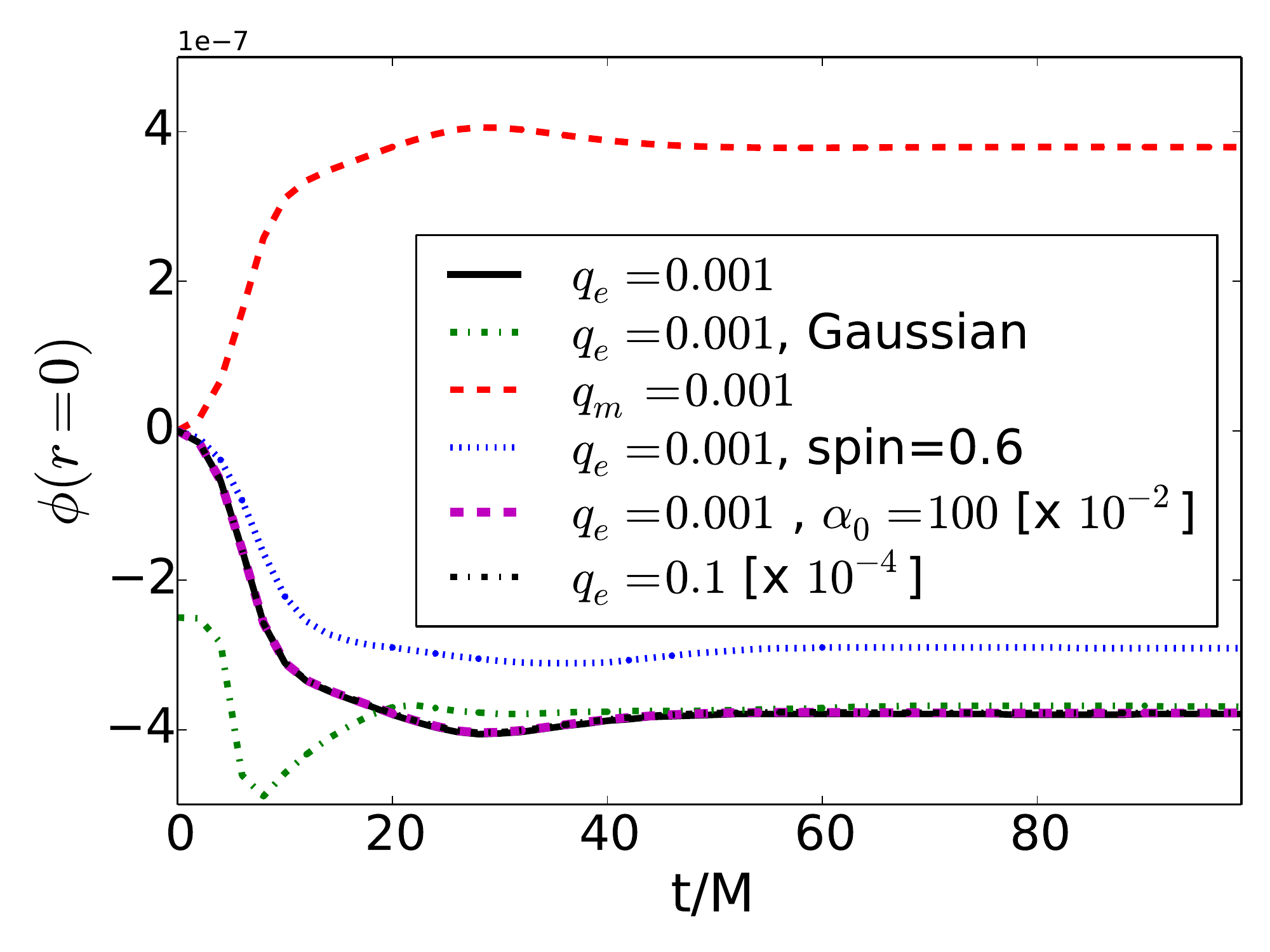}
\caption{Central value of dilaton scalar field as a function of time for various
configurations of a single black hole. 
Note that the system generally settles to an apparently stationary
configuration for some non-trivial profile of the dilaton field.
Here, the asymptotic value of the scalar field for the electric cases is
$\phi_{0}=-10^{-10}$ and for the magnetic case $\phi_{0}=+10^{-10}$.
For those not denoted otherwise, $\alpha_0=1$.
The last two cases are rescaled as indicated to fit the scales of the plot.
} 
\label{fig:bh_phimin}
\end{figure}

This initial data is evolved and, after some transient behavior, 
the system generally settles into a stationary solution.
This behavior can be appreciated
in Fig.~\ref{fig:bh_phimin} which plots the central value of $\phi$ as a function
of time for a number of different configurations.
In particular, for small values of $\alpha_0$, as the charge $q \equiv Q/M$ is increased, 
the magnitude of the dilaton increases quadratically (the figure rescales some of the curves to fit in the figure); 
similar behavior is observed when increasing the value of $\alpha_0$, which induces a linear increase
in the dilaton.  This behavior is in agreement with the known solution described in Appendix~\ref{knownsolutions}.

Also shown in the figure are the results of choosing a Gaussian profile for
the dilaton at the initial time instead of a constant value. In particular, adopting
a Gaussian centered at the origin results in essentially the same stationary solution, despite
differences at early times. This agreement suggests that a unique, static hairy black hole,
insensitive to the initial configuration of the dilaton, is an attractor.

Also included is one example of a spinning black hole with $a/M = 0.6$. Similar to
the static case, our evolutions suggest a unique, stationary, rotating, stable  hairy
black hole. The effect of increasing $\alpha_0$ and the spin are discussed below.

Finally, there is one case where a monopole
magnetic field has been added to the black hole,
showing that the dilaton is basically the same
as in the electric case but with the opposite sign.

To analyze the solution in more detail, we focus on the case of a non-spinning, 
electrically charged black hole with $q_e\equiv Q_e/M=10^{-3}$  and obtain the 
asymptotic dependence of the field which can be described by
$\phi(r) \approx \phi_0 + \phi_1/r$. 
We stress that, in contrast to the behavior of the scalar field with neutron 
stars in ST theories, the scalar charge $\phi_1$ does not sensitively depend 
on the 
asymptotic value of the scalar field $\phi_0$. This insensitivity implies that
effects such as  induced and dynamical 
scalarization
are less significant in EMD than in ST theories~\cite{Barausse:2012da,Palenzuela:2013hsa}.

In relation to the discussion in Appendix~\ref{instab}, Fig.~\ref{fig:charges} 
shows the value of this scalar charge as a function of dilaton coupling 
$\alpha_0$. Notice the linear behavior for small values of $\alpha_0 $ 
whereas a different trend is clear for larger values. This behavior can be 
extracted analytically from the solution presented in 
Appendix~\ref{knownsolutions} (neglecting, for the moment, the asymptotic 
value of the dilaton) and from which the scalar charge can be calculated as
\begin{equation}
\phi_1 = {\alpha_0 Q_e^2 \over M} \, {1 \over 1 + \sqrt{ 1 + (\alpha_0^2 -1) Q_e^2/M^2} } \, .\label{expectedcharge}
\end{equation}
The behavior at small $\alpha_0$ extracted from Eq.~(\ref{expectedcharge}) is 
$\phi_1 \approx \alpha_0 Q_e^2/(2 M)$ 
while for large values $\phi_1 \rightarrow |Q_e|$. The numerical
solutions obtained for $\alpha_0 \lesssim 5000$ are in excellent agreement with this expression while
a lower than expected scalar charge is obtained above this value of $\alpha_0$. We note however that
numerical simulations become quite challenging at such large values.
For this reason, we will restrict to $\alpha_0 \le 3000$ when studying binary mergers\footnote{
Up to this value of the coupling, the analytic and late-time numerical 
solutions for the single BH agree.
}.

We also monitor the central value of the scalar field and display
its behavior for the electric case in the bottom panel of Fig.~\ref{fig:charges}.
Although the central field increases at small coupling, the trend changes
dramatically around $\alpha_0=2000$, precisely the point at which  the dilaton charge 
saturates. This behavior is yet another indication of a transition in the system
as $\alpha_0$ is increased.

\begin{figure}[h]
\centering
\includegraphics[width=8.2cm,angle=0]{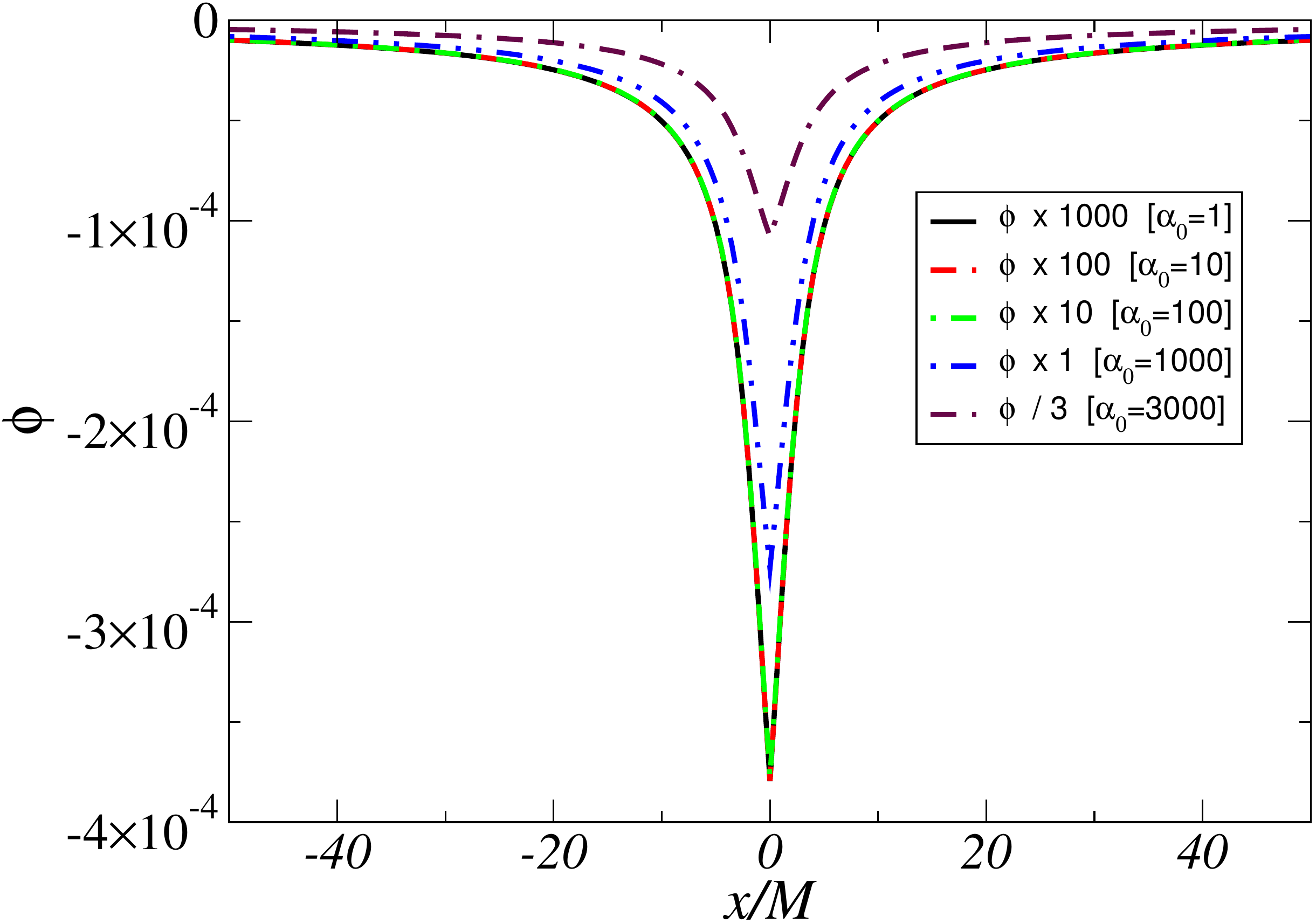}
\caption{The profile of the scalar field for different values of $\alpha_0$ (with $q_e=0.001, M=1$ and $\phi_{0}=10^{-10}$).
The profiles are rescaled assuming a linear increase with the coupling. 
For $\alpha_0 \lessapprox 100$ 
this linear  scaling is apparent.
} 
\label{fig:phivsr}
\end{figure}

Additional insights can be gained by examining the radial profile of the stationary solution and its dependence on the coupling $\alpha_0$.
Fig.~\ref{fig:phivsr} shows the radial profile of the dilaton in the case that $\phi_0 = 10^{-10}, q_e = 10^{-3}$ and for  
$\alpha_0=\{1, 10, 10^2, 10^3, 3 \times 10^3\}$. For comparison purposes, we rescale linearly the profiles with respect to
the value $\alpha_0=1000$. 
That the solution scales linearly for small coupling is clearly apparent,
in contrast to the solutions for large $\alpha_0$.

\begin{figure}[h]
\centering
\includegraphics[width=8.2cm,angle=0]{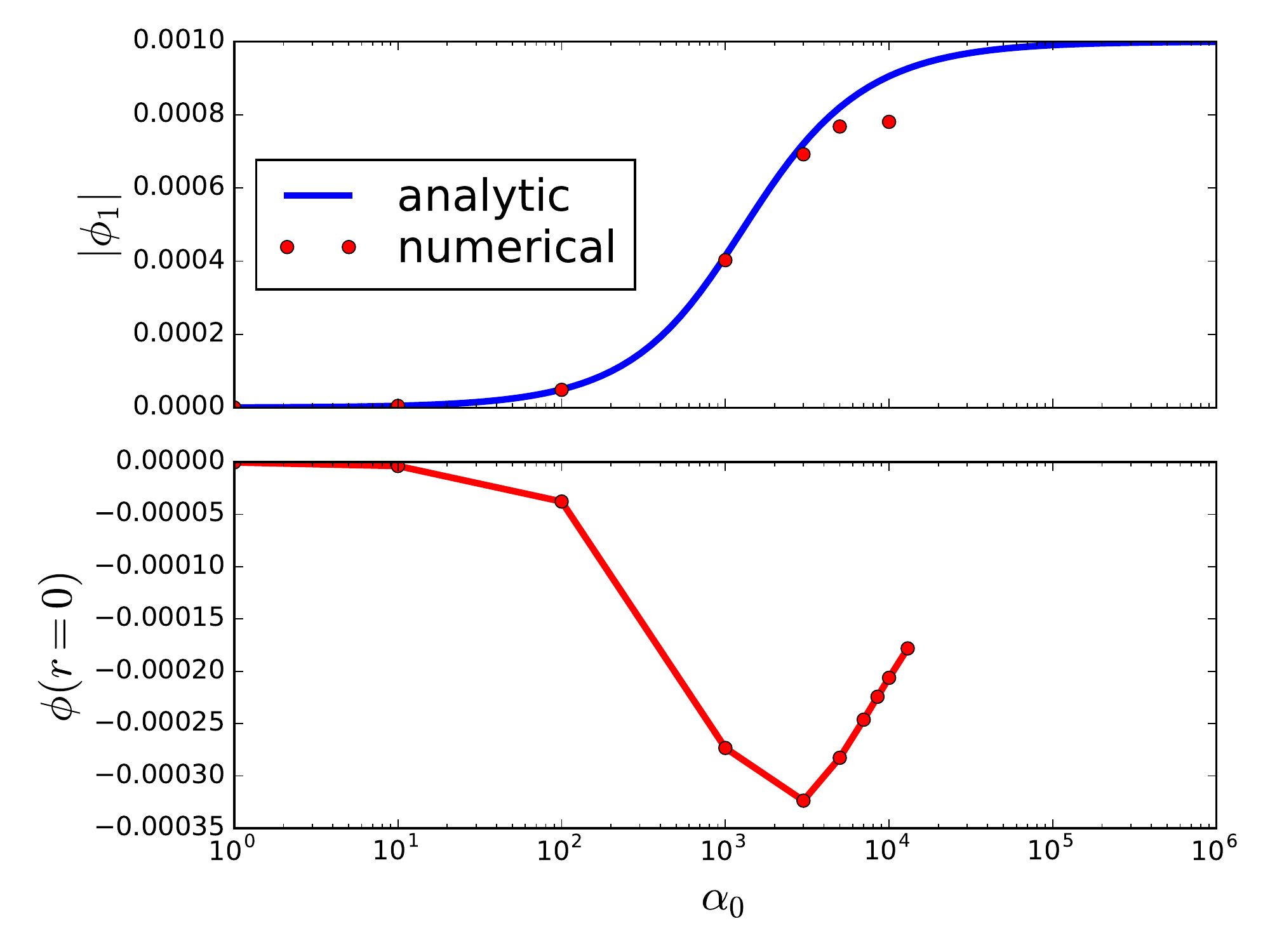}
\caption{Dilaton behavior for a single, electrically charged, non-spinning 
black hole
with $q_e=0.001$, $M=1$, and $\phi_{0}=-10^{-10}$. These quantities
have been extracted at late times when the solution settled down to
a roughly stationary solution.
(Top) The dilaton scalar charge along with the analytic value, expression
~(\ref{expectedcharge}), for $q_e=0.001$ and $M=1$.
(Bottom) The central value of the scalar field at late times.
} 
\label{fig:charges}
\end{figure}

Proceeding in a similar fashion for spinning black holes (keeping for concreteness $q_e=0.001, \alpha_0=1$),
we measured the scalar charge as we vary the spin parameter of the black hole  in the range $[0,0.6]$ 
(to ensure a small initial constraint violation). We find
that the scalar charge measured from the stationary state can be fit approximately as
\begin{equation}
   \phi_1 (a/M) = \phi_1 (0) 
    \left(1 - 0.4 \left(\frac{a}{M}\right)^{2}\right) 
\end{equation} 
where $\phi_1(a=0)$ is the value of the scalar  charge for the non-rotating case. Notice, the charge
{\em decreases} as the spin increases. 

We have also looked at the quasi-normal modes (QNM) of oscillation
of perturbed black holes. In particular, on simulating the head-on collision 
of two black holes we produce a strongly perturbed remnant black hole
and can extract the frequency of the strongest QNM in the 
ringdown.  Such QNM frequencies have been  
calculated analytically for $\alpha_0=1$ in EMD~\cite{Ferrari:2000ep}.
We confirm that the frequency from the numerical simulation agrees with
the analytic value to within $4(7)\%$ for the real (imaginary) part of the
frequency 

However, we note that for the small (EM) charges that we consider here, the
differences in these QNM frequencies in comparison to the GR
case is small not just for the $\alpha_0=1$ case, but for a large range of
$\alpha_0$ values as well.
In consequence, the difference in EMD and GR ringdown dynamics will not be
distinguishable above our numerical error (of the order of $5\%$ in the
extracted frequency/decay rate of the fundamental mode).  Another way of saying this is
that the role of the dilaton is largely inconsequential in terms of its
effect on the dynamics and the formation of the final black hole.
This is consistent with our earlier observation that EMD, for different
values of the coupling $\alpha_0$, has a phenomenology that interpolates
between charged and neutral black holes.  Indeed, we could well have inferred
the comparable QNM decay rates between EMD and GR from this observation and
the known QNM spectra for weakly charged black
holes~\cite{PhysRevLett.110.241103,PhysRevD.90.124088,Mark:2014aja,Dias:2015wqa}.

\subsection{Binary black holes}
We now turn our attention to binary black hole systems both with equal and unequal masses. From our single
black hole results, it is clear that observations made with small values of the coupling $\alpha_0$
have a simple scaling until $\alpha_0 \approx 10^3$ for the cases where the charge is $q_e=10^{-3}$. We have
studied the dynamics of binaries for a broad set of $\alpha_0$ values and confirmed this expectation.
In what follows we thus concentrate on discussing the particular cases $\alpha_0 = \{1, 10^3, 3 \times 10^3\}$.

\begin{figure}[h]
\centering
\includegraphics[width=8.8cm,angle=0]{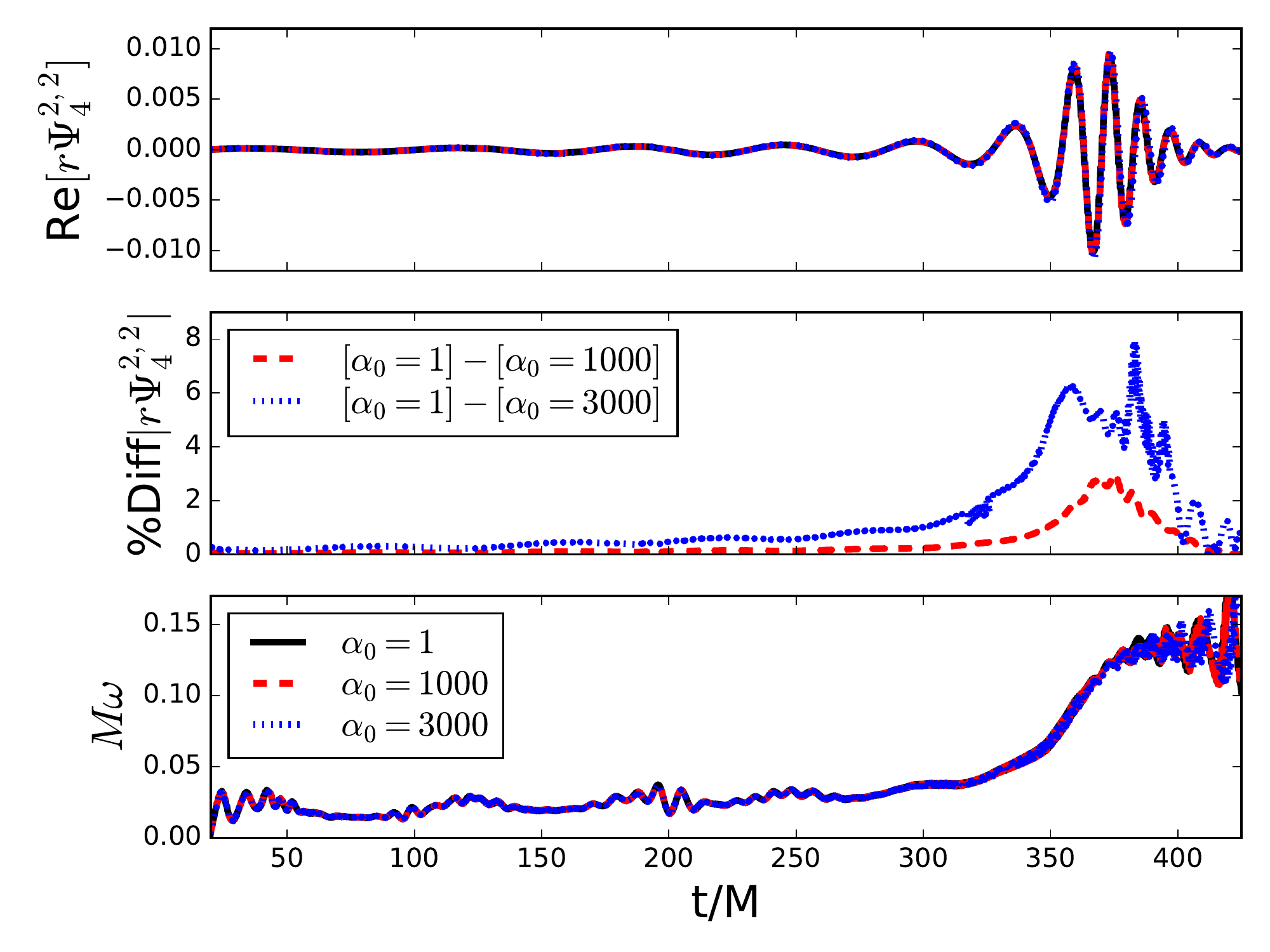}
\caption{ The gravitational radiation of a {\bf equal mass} binary black hole with an electric charge $q_e = 0.001$ for different values of $\alpha_0$. 
\textbf{Top:} The real part of the $l=m=2$ mode of the Newman-Penrose scalar $\Psi_4$.
\textbf{Middle:} The percent difference in magnitude of the $l=m=2$ mode of $\Psi_4$
relative to the $\alpha_1$ case normalized by the maximum of the signal.
\textbf{Bottom:} The angular frequency of the gravitational wave mode.}
\label{fig:trajectories_alpha0}
\end{figure}

\begin{figure}[h]
\centering
\includegraphics[width=8.8cm,angle=0]{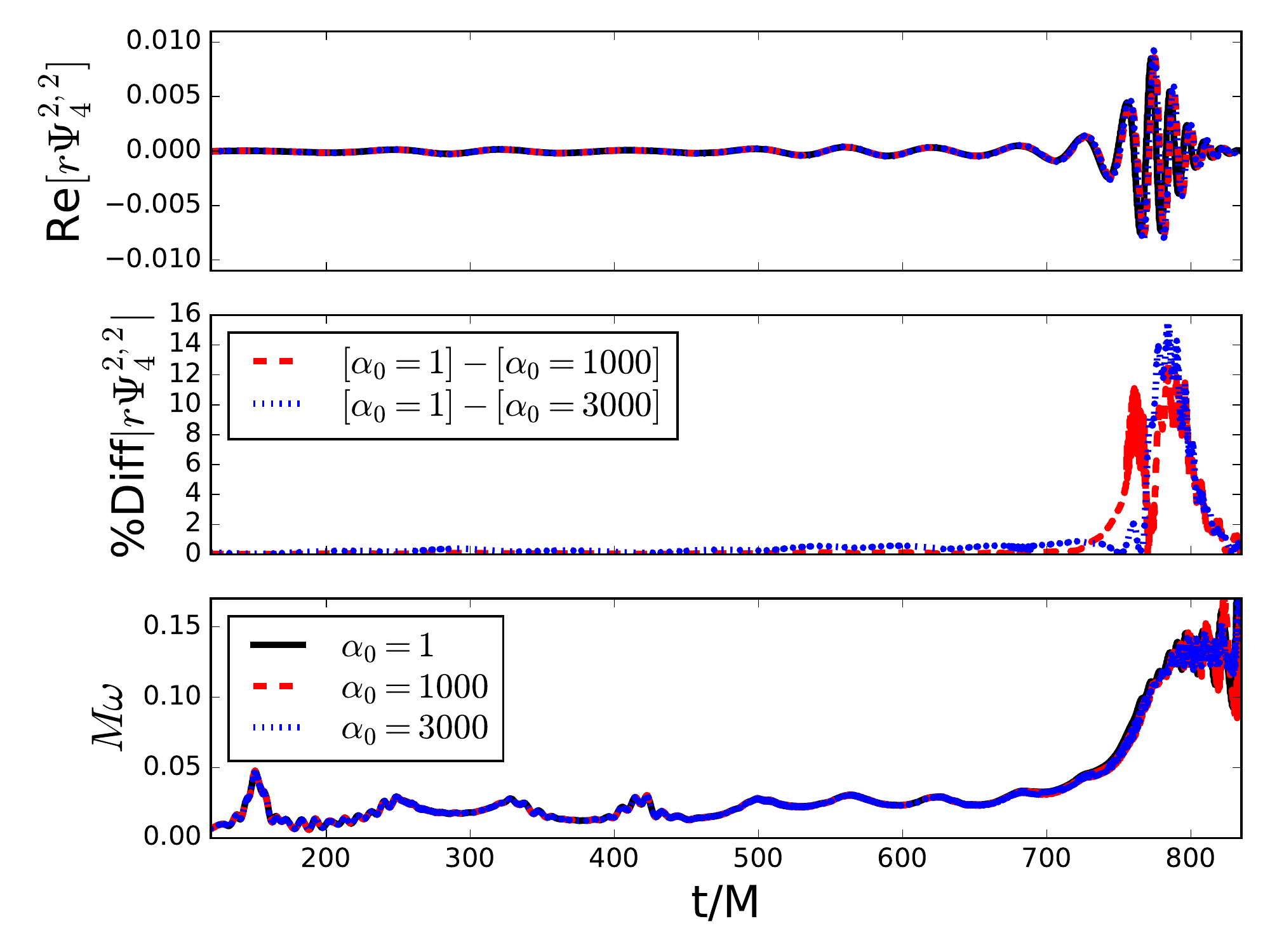}
\caption{ The gravitational radiation of a {\bf unequal mass} binary black hole with an electric charge $q_e = 0.001$ for different values of $\alpha_0$. 
\textbf{Top:} The real part of the $l=m=2$ mode of the Newman-Penrose scalar $\Psi_4$.
\textbf{Middle:} The percent difference in magnitude of the $l=m=2$ mode of $\Psi_4$
relative to the $\alpha_1$ case normalized by the maximum of the signal.
\textbf{Bottom:} The angular frequency of the  gravitational wave mode.
} 
\label{fig:trajectories_alpha0_uneq}
\end{figure}

For the nondimensionalized electric charge of $q_e=10^{-3}$, the binaries orbit for 4-5 cycles before merging into a single spinning black hole.
Figs.~\ref{fig:trajectories_alpha0} and~\ref{fig:trajectories_alpha0_uneq} summarize our results for both equal and unequal mass (mass ratio $m_1/m_2 = 3/2$) cases.
In each figure, the top panel shows the real part of the radiative Newman-Penrose scalar $\Psi_4$ for different values of $\alpha_0$, and the middle panel displays the differences of their magnitudes with respect to the $\alpha_0=1$ case.
These plots show that differences are very small even for $\alpha_0\gg1$. 
The angular frequency of the dominant gravitational wave mode
is shown in the bottom panel. 
Again, the differences with variations in $\alpha_0$ are small. However,
that the cases with larger coupling merge earlier (albeit very slightly)
is consistent with the expectations that larger coupling will result in
increased energy loss through scalar radiation.

Of course even these small differences are possibly degenerate
with other parameters. That is to say that the signals we find, were
they to be measured by LIGO, would likely be mistaken for GR signals for a BH
binary with parameters somewhat different than those we adopt here.
The binary mass in particular could probably be adjusted to generate
GR waveforms that would be indistinguishable from these EMD waveforms.
Recent examples proposing that LIGO's detections are perhaps more exotic
than simply binary black hole mergers within GR include~\cite{Moffat:2016gkd,Cardoso:2016olt}.

The observations that differences are small can also be inferred from multiple points of view, to wit:\\

\vspace{0.15in}

$\bullet$ As noted in Appendix~\ref{knownsolutions}, static black holes in EMD 
become neutral in the $\alpha_0 \rightarrow \infty$ limit.
It is natural then to expect a similar limit for black holes in binaries,
and this decreasing charge has implications for merger time.
As demonstrated in Ref.~\cite{putarakinprep}  
which studied the particular case of black hole binaries with the same sign of charge
with $\alpha_0=\sqrt{3}$ 
(following~\cite{Buonanno:2007sv}), the estimated radius of the effective innermost stable circular orbit~(ISCO) 
increases as the black hole charge  increases.
Following our observation that larger couplings have effectively smaller black hole charges, one
expects that, for fixed charges of equal sign, black holes 
will merge sooner (i.e. at lower frequencies) for smaller coupling.\\

$\bullet$  It is interesting to consider the behavior of binary neutron
stars in scalar tensor theories. In particular, the clearest differences
in those simulations from those of GR occurred for scalar charges of
the order $\phi_1 \approx 10^{-1}$. However, here the charges are a couple orders
of magnitude smaller, only $\phi_1 \approx 10^{-3}$, and 
one expects dynamical differences to scale with $\phi_1^2$.
And so perhaps it is natural that the differences we see for these
parameter choices are as small as we report.
Scalarization levels comparable to those neutron star mergers
would require BH charges $\alpha_0 q^2 M \approx 10^{-1}$ ($q M \simeq 10^{-1}$) 
for small (very large) values of $\alpha_0$.

\vspace{0.15in}

\begin{figure}[h]
\centering
\includegraphics[width=8.8cm,angle=0]{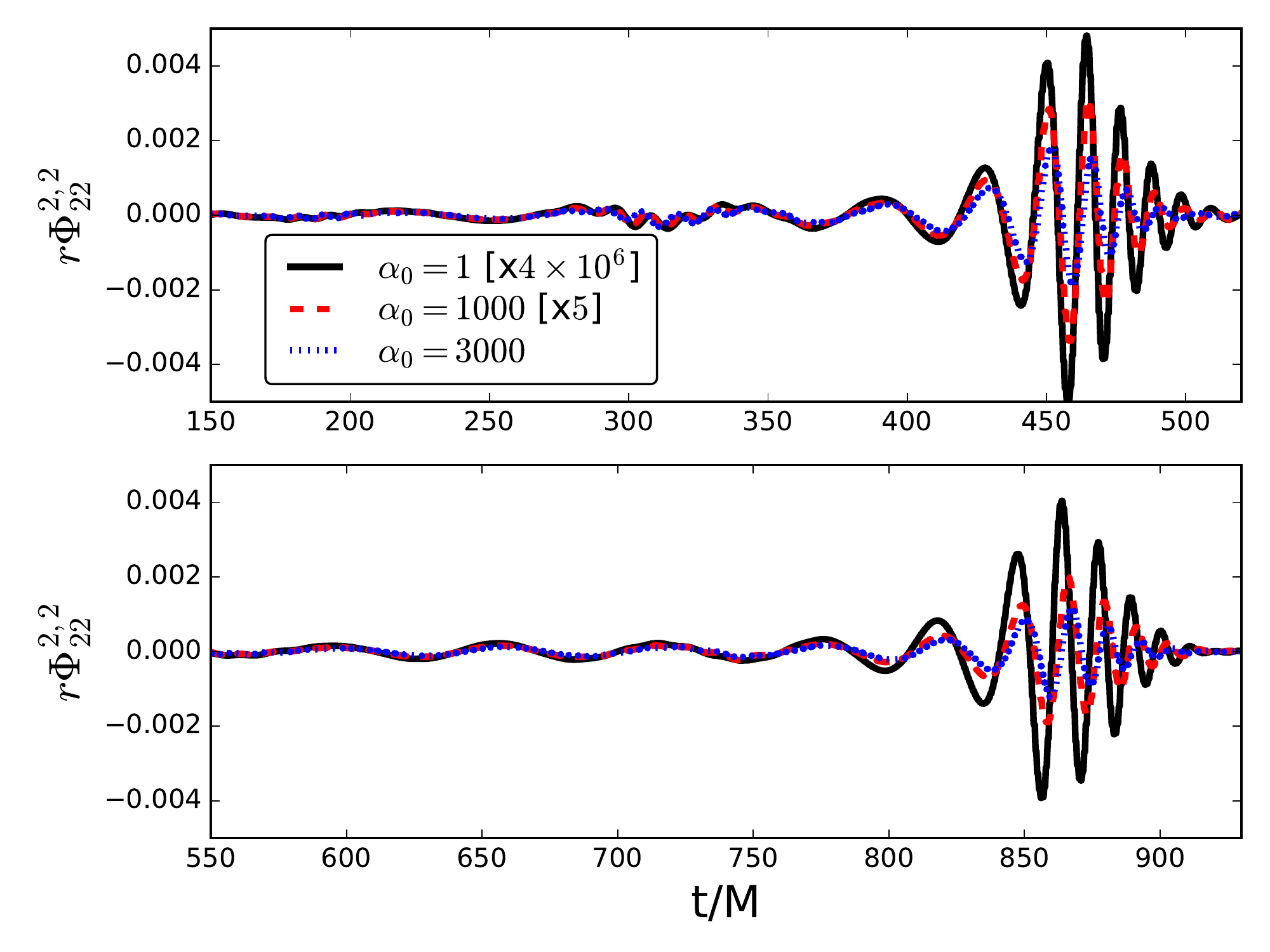}
\caption{ The (real part of the) $l=m=2$ mode  of the scalar gravitational 
radiation $\Phi_{22}$ of a binary black hole with an electric charge $q_e = 0.001$ for different values of $\alpha_0$. 
\textbf{Top:} The equal mass case.
\textbf{Bottom:} The unequal mass case. 
} 
\label{fig:phi2_alpha0}
\end{figure}

Because EMD allows for scalar radiation, we can gain additional understanding by extracting it in addition to the
gravitational wave signal.
As discussed in Appendix~\ref{app:radiation}, 
 the computation of the Newman-Penrose scalar $\Phi_{22}$ indicates that the scalar radiation is expected to scale as
$\Phi_{22} \approx \alpha_0 \phi_{,tt}$ (evaluated asymptotically).  
One can thus estimate that this radiation in the early inspiral phase scales as $\Phi_{22} \approx \alpha_0 \phi_1 \Omega^2$.
This scaling is assumed in 
 Fig.~\ref{fig:phi2_alpha0} which shows
$\Phi_{22}$ as a function of time for both the equal and unequal mass cases. 
In particular, because the orbital frequency differs only slightly with
changes to $\alpha_0$, the rescaling depends only on the coupling and
scalar charge.
The coupling value is straightforward, but the black hole charge is chosen
as the charge of individual black holes in isolation. Thus the charges for
equal mass binaries are chosen as: $\phi_1$= $\{-4.8 \times 10^{-7}, -4 \times 10^{-4}, -6.9 \times 10^{-4}\}$
and for unequal mass binaries ($m_1,m_2$):  $\phi_1$= $\{\ (-3,-2)\times 10^{-7}, (-2.4,-1.6) \times 10^{-4}, (-4.2,-2.7)  \times 10^{-4}  \}$ for $\alpha_0 = \{1, 10^{3}, 3 \times 10^3\}$ respectively (which are well approximated by the analytical expression Eq.~\ref{expectedcharge}).

As shown in Fig.~\ref{fig:phi2_alpha0}, reasonably good agreement with the expected scaling 
is obtained during the inspiral phase, but the scaling
overestimates the magnitude of the radiation during the merger. 
The failure of the scaling during the merger indicates that the nonlinear
behavior is less radiative than otherwise 
expected from simple superposition arguments and is consistent with the observations
made in the isolated black hole cases where the scalar charge shows a trend towards saturation at high 
coupling values.

\begin{figure}[h]
\centering
\includegraphics[width=8.8cm,angle=0]{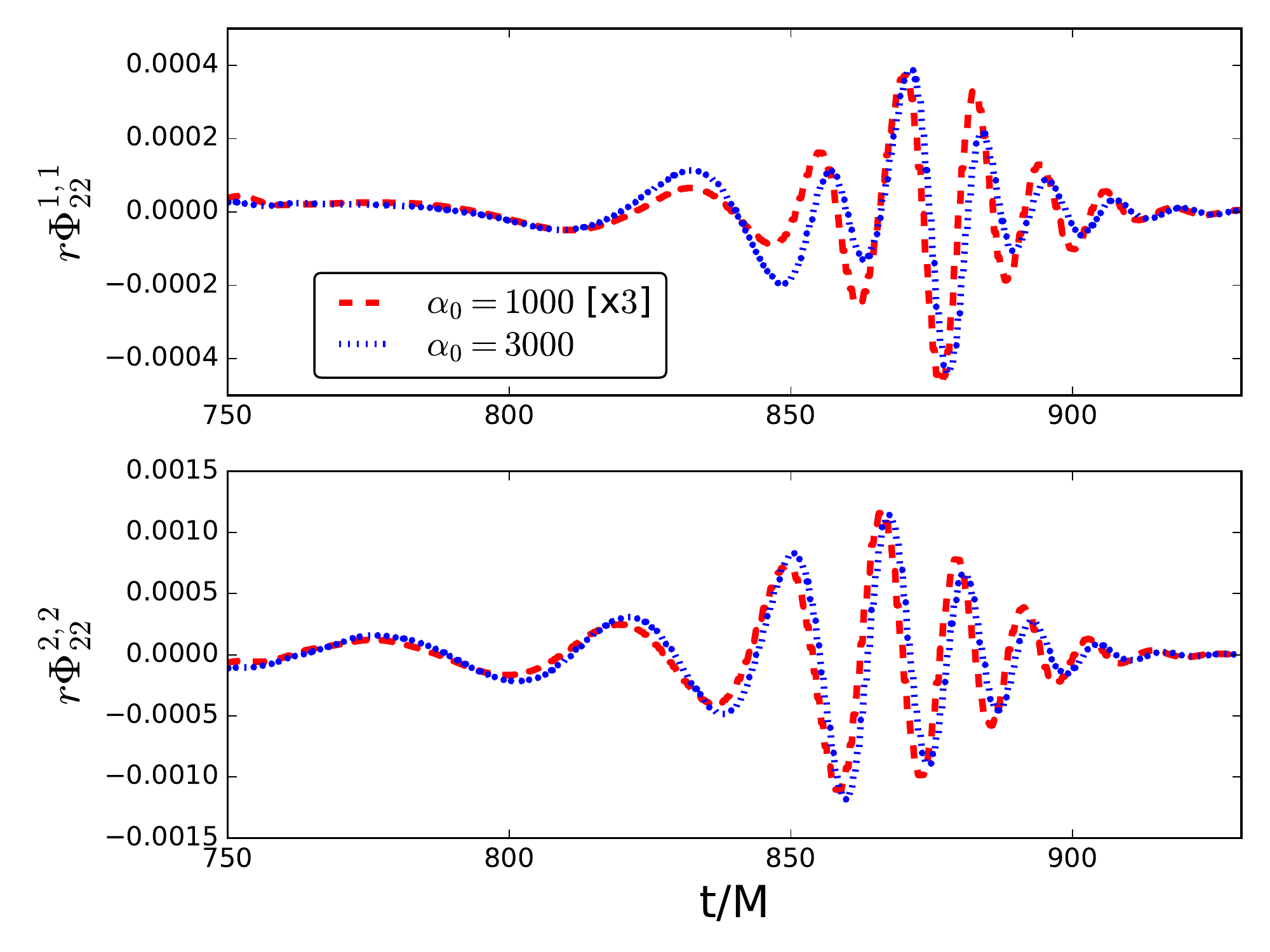}
\caption{ The (real part of the) $l=m=1$ and $l=m=2$ modes  of the scalar gravitational 
radiation $\Phi_{22}$ of a binary black hole with an electric charge $q_e = 0.001$ 
for different values of $\alpha_0$ corresponding to the unequal mass binary case. Here we have scaled up the case $\alpha_0=1000$ by a factor
of $3$ in accordance with  expected scaling if the dilaton charge is ignored.
\textbf{Top:} The $l=m=1$ mode.
\textbf{Bottom:} The $l=m=2$ mode.
} 
\label{fig:phi2_uneq_alpha0}
\end{figure}

This saturation  is evident in Fig.~\ref{fig:phi2_uneq_alpha0} which shows the $l=m=1$ and $l=m=2$ modes
for the scalar radiation corresponding to the unequal mass binary for $\alpha_0=1000$ and $3000$. In contrast with Fig.~\ref{fig:phi2_alpha0} however, both cases
here are scaled linearly by their respective value of $\alpha_0$, ignoring
the dilaton charge. Focusing only on the merger, this simple scaling in
$\alpha_0$ works quite well, supporting our assertion that the scalar charges
saturate at large coupling. 

An interesting aspect of gravitational radiation in EMD is that it could contain a dipolar component 
in contrast to GR which disallows dipolar radiation.
Although one generally expects the dipolar component, when allowed by the theory, to dominate over higher multipoles,
here the strength of the dipolar component depends on the difference in the scalar charges of the black holes.
As a result,  the equal mass case produces no dipolar radiation.
For the unequal mass binary with $m_1/m_2=3/2$ considered here, the scalar charges
are different, but nevertheless are
sufficiently close to each other that the resulting $l=m=1$ mode is weaker than the $l=m=2$ mode.

We comment on two further conclusions that can be drawn from our studies as well as leave open a question deserving of investigation.
First, we find the ringdown of the merger remnant appears largely insensitive to the value of the coupling. As mentioned
in the previous section concerning the ringdown of the head-on remnant, 
small values of the electric charge produce correspondingly small differences in ringdown versus GR.

Second, we also studied the merger of black holes with electric charges of opposite sign.
For the small electric charges
considered here, no significant effect was observed on the black hole dynamics, indicating, as one might expect, 
that electromagnetic forces are sub-leading with respect to gravitational ones. 

Finally, recall that the dilaton permits EMD black holes to have electric 
or magnetic charge (or both) and, as a result, these black holes have different 
properties.  It could be interesting to consider the interaction of a binary
black hole system comprised of one electrically charged and one magnetically 
charged black hole and investigate the impact on the dynamics and 
radiation.   
In particular, the black holes in such a binary would have scalar
charges of opposite sign which might maximize the resulting dipole scalar radiation.
However, a preliminary investigation with the small charge used here
did not reveal any dramatic effects.

\section{Final discussion}\label{finaldiscussion}
We have examined the dynamics of black holes, both in isolation and in
binaries,
within Einstein-Maxwell-dilaton theory. We have 
focused on the differences between these dynamics and those in general
relativity.

This theory is parametrized by a coupling constant, $\alpha_0$. For $\alpha_0=0$,
the theory describes Einstein-Maxwell with a free scalar field, and its black hole
solutions include Reissner-Nordstrom. The low energy limit of string theory
is described by $\alpha_0=1$ which includes hairy black hole solutions.
In the infinite coupling limit $\alpha_0 \rightarrow \infty$, 
the single, spherically symmetric, black hole solution is simply the 
Schwarzschild solution of pure vacuum general 
relativity 
and the electromagnetic field is essentially ``screened'' out.  

Our results for binary mergers appear consistent with these same limits for isolated
black holes. Of course for $\alpha_0=0$ our black holes merge producing a charged,
hairless black hole. As $\alpha_0$ is increased, the remnant black hole displays
a scalar charge (in addition to its angular momentum and Maxwell charge).

The dynamics of both the dilaton and the gauge
field are important and can impact the behavior of the dynamics of the binary. 
This influence is primarily governed by the strength of the scalar charge
of the black hole which scales, at small coupling, as $\alpha_0 Q^2$.   
At large coupling, however, the scaling is such that the dilaton charge 
scales linearly with $Q$.  For small values of $Q$, as we have seen, the 
effects are minor while we expect 
large effects
for larger values of $Q$.

Interestingly, because the scalar charge in EMD does not depend sensitively 
on the asymptotic value of the dilaton or the nearby charge of a companion
(as opposed to the case in scalar tensor theory~\cite{PhysRevLett.70.2220,Barausse:2012da}), 
its main role in equal mass binaries can be approximated by charged binary black hole
mergers.

Considering again the case of large coupling, it is worth pointing out that the 
$\alpha_0 \rightarrow \infty$ limit is essentially a decoupling limit  
such that the gravitational dynamics and the matter (Maxwell and dilaton) 
dynamics have decreasing effect on one another.  For large $\alpha_0$, the 
matter fields are increasingly radiated away while the scalar and 
electromagnetic contributions to the final black hole go to zero in this 
limit.  

As discussed, for black holes in EMD, little has been known
with regard to their stability properties, their perturbation spectra for 
arbitrary coupling values, rotating solutions, etc.
Our studies have shown that black holes in EMD have stability properties 
similar to those
in GR. These results extend the analytical studies of~\cite{Ferrari:2000ep} and highlight
the small and subtle differences involved in distinguishing BHs in EMD and GR theories. 

Finally, an immediate conclusion of our work is that for small charges, differences with respect to
waveforms in GR and EMD are quite small. Larger charges may well produce significant differences and
their main characteristics could be bracketed by analyzing charged/uncharged 
collisions in GR~\cite{Zilhao:2012gp,Liebling:2016orx}. While
one does not expect significantly charged black holes in the universe, it is important to stress
that the gauge field in EMD need not be the physical one coupled to the Standard Model.

\begin{acknowledgments}
We thank Thibault Damour, William East, Matt Johnson, Paolo Pani and Frans 
Pretorius as well as our longtime collaborators
Mathew Anderson, David Neilsen and Patrick Motl for interesting discussions.
This work has been supported in part by: NSF grants PHY-1308727 and PHY-1607356~(BYU), PHY-1607291~(LIU), 
NSERC and CIFAR (LL), the Spanish Ministry of Economy and Competitiveness grant AYA2016-80289-P (AEI/FEDER,UE) to CP.
This research was enabled in part by support provided by scinet (www.scinethpc.c
a) and Compute Canada (www.computecanada.ca).
Research at Perimeter
Institute is supported through Industry Canada and by the Province of Ontario
through the Ministry of Research \& Innovation.

\end{acknowledgments}

\appendix

\section{EMD Black hole solutions in isotropic coordinates}\label{app:bhistropic}

Given our use of the BSSN formalism, isotropic coordinates are particularly 
useful.  We present the spherically symmetric, static EMD black hole solutions
here for reference.  Defining a radial, isotropic coordinate, ${\bar r}$,
via
\begin{equation} 
r = {1 \over {\bar r}} \, \biggl[ \Bigl( {\bar r} + {r_{+} + r_{-} \over 4} \Bigr)^2 - {r_{+} r_{-} \over 4} \biggr] 
\end{equation} 
for which 
\begin{eqnarray}
r_{+} & = & M \, \biggl\{ 1 + \Bigl[ 1 - \bigl(1-\alpha_0^2\bigr) \, {Q^2 \over M^2} \Bigr]^{1/2} \, \biggr\} \\
    r_{-} & = & {Q^2 \over M} \, \bigl(1+\alpha_0^2\bigr) \, \biggl\{ 1 + \Bigl[ 1 - \bigl(1-\alpha_0^2\bigr) {Q^2 \over M^2} \, \Bigr]^{1/2} \, \biggr\}^{-1} 
\end{eqnarray}
we can write the metric for both magnetic and electric cases as 
\begin{eqnarray} 
{\rm d}s^2 & = & - \alpha^2 {\rm d}t^2 + \chi^{-1} \bigl[ {\rm d}{\bar r}^2 + {\bar r}^2 \, {\rm d}\Omega^2 \bigr]  \\  
           & = & - { \bigl( {\bar r} - {\bar r}_H \bigr)^2 \bigl( {\bar r} + {\bar r}_H \bigr)^{2(1-\alpha_1)} \over \bigl( {\bar r} + {\bar r}_1 \bigr)^{2-\alpha_1} \bigl( {\bar r} + {\bar r}_2 \bigr)^{2-\alpha_1} } \, {\rm d}t^2 
           \cr 
           & & + {1\over {\bar r}^4} \bigl( {\bar r} + {\bar r}_1 \bigr)^{2-\alpha_1} \bigl( {\bar r} + {\bar r}_2 \bigr)^{2-\alpha_1} \bigl( {\bar r} + {\bar r}_H \bigr)^{2\alpha_1} \, \nonumber \\
& & \times \, \bigl[{\rm d}{\bar r}^2 + {\bar r}^2 \, {\rm d}\Omega^2 \bigr] .
\end{eqnarray}
Here we have defined
\begin{eqnarray}
    {\bar r}_1 & = & {1\over4} \bigl( \sqrt{r_{+}} - \sqrt{r_{-}} \bigr)^2 \\
    {\bar r}_2 & = & {1\over4} \bigl( \sqrt{r_{+}} + \sqrt{r_{-}} \bigr)^2 \\
    {\bar r}_H & = & {1\over4} \bigl( r_{+} - r_{-} \bigr) 
\end{eqnarray}
with ${\bar r}_H$ the radial location of the horizon in these coordinates.
If we consider the magnetic case, then $Q^2 = Q_m^2 e^{-2\alpha_0\phi_0}$ while
for the electric case, $Q^2 = Q_e^2 e^{2\alpha_0\phi_0}$.   

In the magnetic case, the EM and dilaton fields take
the form
\begin{eqnarray}
F_{\theta\phi} & = & Q_m \sin\theta \\
B^{\bar r} & = & { Q_m \, {\bar r}^4 \over \bigl( {\bar r} + {\bar r}_H \bigr)^{3\alpha_1} } \, \Bigl[ \bigl( {\bar r} + {\bar r}_1 \bigr) \bigl( {\bar r} + {\bar r}_2 \bigr) \Bigr]^{3(\alpha_1 - 2)/2} \qquad \\ 
e^{-2\alpha_0 \phi} & = & e^{-2\alpha_0 \phi_0} \, { \bigl( {\bar r} + {\bar r}_H \bigr)^{2\alpha_1} \over \bigl( {\bar r} + {\bar r}_1 \bigr)^{\alpha_1} \bigl( {\bar r} + {\bar r}_2 \bigr)^{\alpha_1} } .
\end{eqnarray}
In the electric case, the EM and dilaton fields take the form
\begin{eqnarray}
F_{t{\bar r}} & = & Q_e \, { \bigl( {\bar r}^2 - {\bar r}_H^2 \bigr) \over \bigl( {\bar r} + {\bar r}_1 \bigr)^2 \bigl( {\bar r} + {\bar r}_2 \bigr)^2 } \\
E^{\bar r} & = & - { Q_e \, {\bar r}^4 \over \bigl( {\bar r} + {\bar r}_H \bigr)^{\alpha_1} } \, \Bigl[ \bigl( {\bar r} + {\bar r}_1 \bigr) \bigl( {\bar r} + {\bar r}_2 \bigr) \Bigr]^{(\alpha_1-6)/2} \qquad \\ 
e^{2\alpha_0 \phi} & = & e^{2\alpha_0 \phi_0} \, { \bigl( {\bar r} + {\bar r}_H \bigr)^{2\alpha_1} \over \bigl( {\bar r} + {\bar r}_1 \bigr)^{\alpha_1} \bigl( {\bar r} + {\bar r}_2 \bigr)^{\alpha_1} }  .
\end{eqnarray}

\begin{figure}[h]
\centering
\includegraphics[width=8.2cm,angle=0]{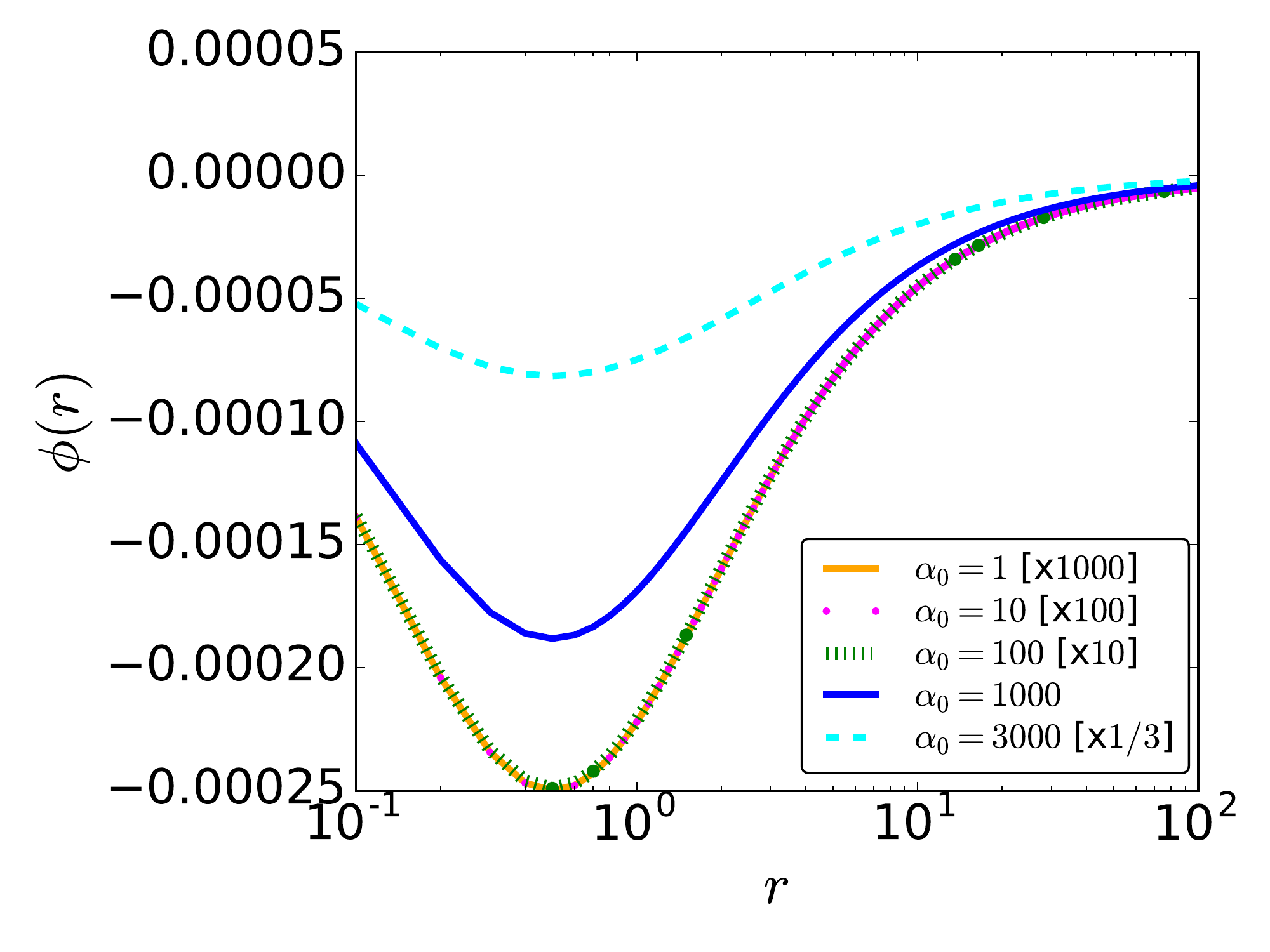}
\caption{Radial profile for the scalar field obtained in isotropic coordinates with different values of $\alpha_0$ (all with
$q_e=10^{-3},\phi_0=10^{-10}$). The solutions have been rescaled in the same
way as the solutions in Fig.~\ref{fig:phivsr}. This linear scaling holds
only up to roughly $\alpha_0 \approx 100$.
}
\label{fig:istropicphi}
\end{figure}

As an illustration of the properties of the solution and a demonstration
of the scaling with the coupling $\alpha_0$, we plot the radial profile of $\phi$ versus
radius 
for different values of $\alpha_0=\{1,10^{1},10^{2},10^{3},3\times 10^3\}$
with fixed $q_e=10^{-3},\phi_0 = 10^{-10}$.
To simplify the comparison, we scale all profiles linearly in $\alpha_0$. 
As shown in Fig.~\ref{fig:istropicphi}, that the rescaled profiles for small $\alpha_0$ coincide demonstrates that the solution
does scale linearly in $\alpha_0$, but the dependence of the solution on $\alpha_0$ is milder at larger values.

The charge of these black holes is given by Eq.~\ref{expectedcharge} and is plotted
in Fig.~\ref{fig:charges} along with the charge obtained in our 3D evolutions (in
different coordinates). The figure makes clear that the charge saturates at large
coupling.

\section{Calculating radiative properties of the solution}\label{app:radiation}
We recall that the physical frame is the Jordan one (the one with respect 
to which particles travel along geodesics). However, in our numerical
studies we find it convenient to compute the evolution in the Einstein frame.
It is thus important to compute the radiative behavior in the Jordan frame
which, in particular, facilitates the comparison across the different cases considered.

Let us then analyze what a Jordan-frame observer would measure
with respect to the  Newman-Penrose radiative scalars obtained in the Einstein frame.
First, recall our conformal transformation from the Jordan to Einstein frame
\begin{equation}
g_{ab}^E = g_{ab}^J e^{-2 \alpha_0 \phi} \equiv g_{ab}^J \varPhi
\end{equation}
where we introduce as a shorthand $\varPhi \equiv e^{-2 \alpha_0 \phi}$. 
Next, given the standard null tetrad chosen in the Einstein frame $T^E_{\alpha}$
(with $\alpha=0..3$ labeling the different null vectors of the tetrad), the Jordan
frame tetrad is trivially related to the Einstein one by
\begin{equation}
T^E_{\alpha} = T^J_{\alpha} \sqrt{\varPhi}.
\end{equation}
Now to find the radiative (spin-2) scalar  $\Psi_4$ in the Jordan frame computed from the Weyl tensor, we
exploit the fact that the Weyl tensor, $C_{abc}^d$, is invariant with respect to
conformal transformations; therefore $C^E_{abcd} = C^J_{abcd} \varPhi$ and,
since $\Psi_4$ involves contractions with $4$ tetrad
members, we have --schematically-- 
\begin{eqnarray}
\Psi^E &=& C^E_{abcd} T^E  T^E  T^E  T^E  \nonumber  \\
       &=& C^J_{abcd} \varPhi \, T^J  T^J  T^J  T^J  (\varPhi)^{-2} \nonumber  \\
       &=& \Psi^J  (\varPhi)^{-1}.
\end{eqnarray}
Thus, $\Psi^J = \Psi^E e^{-2 \alpha_0 \phi}$.

We turn our attention now to the (spin-0) scalar radiation which
in the Newman-Penrose formalism is represented by the real scalar $\Phi_{22}$
and is obtained from the Riemann tensor. 
Recall that
this tensor transforms under conformal transformations as
\begin{equation}
R^E_{ab} = R^J_{ab} - 2 \nabla_a \nabla_b \ln \varPhi + 2 \nabla_a \ln \varPhi \nabla_b \ln \varPhi 
+ g_{ab} {\cal S}
\end{equation}
where ${\cal S}$ contains derivatives of $\varPhi$ but will not contribute since 
$\Phi_{22} \equiv R_{ab} n^a n^b/2$ and $n^a$ is a null vector (the same appearing in
the calculation of $\Psi_4$). 
Proceeding as before, we obtain
\begin{eqnarray}
\Phi^E_{22} &=& R^E_{ab} T^E  T^E/2 \nonumber \\
&=& R^E_{ab} T^J  T^J/2 (\varPhi)^{-1} \nonumber \\
&=& (\Phi^J_{22} - n^a_J n^b_J \nabla_a \nabla_b \ln \varPhi + ...) (\varPhi)^{-1}
\end{eqnarray}
where we denote with $...$ terms proportional to $(\ln \varPhi)^2$ which will be subleading. 
Consequently, we have
\begin{equation}
\Phi^J_{22} = e^{-2 \alpha_0 \phi} \left(\Phi^E_{22} -2 \alpha_0 n^a_E n^b_E \nabla_a \nabla_b \phi  \right)\label{phi22JE}.
\end{equation}
To estimate $\Phi^E_{22}$ we can make use of the trace-reversed form of the Einstein equations 
in the Einstein frame (in what follows we restrict to the Einstein frame and 
we do not include a subindex) to obtain
\begin{equation}
\Phi^E_{22} = R_{ab} n^a n^b/2 = n^a n^b \left(\nabla_a \phi \nabla_b \phi +
2 e^{-2 \alpha_0 \phi} F_{ac} F^c_b  \right)
\end{equation}
where we have dropped terms involving $n^a n^b g_{ab}$. Furthermore, $n^a n^b  F_{ac} F^c_b \propto r^{-2}$
as it can be written in terms of the Newman-Penrose scalar $\phi_2 \bar \phi_2$.
Therefore the contribution of the Einstein-frame $\Phi_{22}$  is subleading with respect to the 
second term in the righthandside of Eq.~(\ref{phi22JE}). To leading
order then, the scalar radiation, measured in the Jordan frame, scales as
\begin{equation}
\Phi_{22}^J \simeq \alpha_0 \phi_{,tt} e^{-2 \alpha_0 \phi}.
\end{equation}
As we have seen, for small values of the coupling $\alpha_0$ the magnitude of the scalar
charge $\phi_1$ grows but such growth saturates, and then reverses at $\alpha_0 \simeq 3000$.

As a last step, one should be mindful of whether the asymptotic time measured in the different
frames coincide. In all our simulations we have chosen the asymptotic value of the scalar field
to be (a small) constant $\phi_{0}$. Upon transformation to the Jordan frame, this implies
asymptotic observers carry clocks ticking at different rates given by $\kappa \equiv e^{\alpha_0 \phi_{0}}$. Thus,
we perform one last transformation to a single, common time, defined by $t \equiv \int  \kappa dt'$ but, for 
the couplings considered and the value of $\phi_{0}=10^{-10}$ adopted here, the correction is negligible.




\providecommand{\href}[2]{#2}\begingroup\raggedright\endgroup

\end{document}

%% file: aas_macros.tex
\def\jnl@style{\it}
\def\aaref@jnl#1{{\jnl@style#1}}

\def\aaref@jnl#1{{\jnl@style#1}}

\def\aj{\aaref@jnl{AJ}}                   
\def\araa{\aaref@jnl{ARA\&A}}             
\def\apj{\aaref@jnl{ApJ}}                 
\def\apjl{\aaref@jnl{ApJ}}                
\def\apjs{\aaref@jnl{ApJS}}               
\def\ao{\aaref@jnl{Appl.~Opt.}}           
\def\apss{\aaref@jnl{Ap\&SS}}             
\def\aap{\aaref@jnl{A\&A}}                
\def\aapr{\aaref@jnl{A\&A~Rev.}}          
\def\aaps{\aaref@jnl{A\&AS}}              
\def\azh{\aaref@jnl{AZh}}                 
\def\baas{\aaref@jnl{BAAS}}               
\def\jrasc{\aaref@jnl{JRASC}}             
\def\memras{\aaref@jnl{MmRAS}}            
\def\mnras{\aaref@jnl{MNRAS}}             
\def\pra{\aaref@jnl{Phys.~Rev.~A}}        
\def\prb{\aaref@jnl{Phys.~Rev.~B}}        
\def\prc{\aaref@jnl{Phys.~Rev.~C}}        
\def\prd{\aaref@jnl{Phys.~Rev.~D}}        
\def\pre{\aaref@jnl{Phys.~Rev.~E}}        
\def\prl{\aaref@jnl{Phys.~Rev.~Lett.}}    
\def\pasp{\aaref@jnl{PASP}}               
\def\pasj{\aaref@jnl{PASJ}}               
\def\qjras{\aaref@jnl{QJRAS}}             
\def\skytel{\aaref@jnl{S\&T}}             
\def\solphys{\aaref@jnl{Sol.~Phys.}}      
\def\sovast{\aaref@jnl{Soviet~Ast.}}      
\def\ssr{\aaref@jnl{Space~Sci.~Rev.}}     
\def\zap{\aaref@jnl{ZAp}}                 
\def\nat{\aaref@jnl{Nature}}              
\def\iaucirc{\aaref@jnl{IAU~Circ.}}       
\def\aplett{\aaref@jnl{Astrophys.~Lett.}} 
\def\apspr{\aaref@jnl{Astrophys.~Space~Phys.~Res.}}
\def\bain{\aaref@jnl{Bull.~Astron.~Inst.~Netherlands}} 
\def\fcp{\aaref@jnl{Fund.~Cosmic~Phys.}}  
\def\gca{\aaref@jnl{Geochim.~Cosmochim.~Acta}}   
\def\grl{\aaref@jnl{Geophys.~Res.~Lett.}} 
\def\jcp{\aaref@jnl{J.~Chem.~Phys.}}      
\def\jgr{\aaref@jnl{J.~Geophys.~Res.}}    
\def\jqsrt{\aaref@jnl{J.~Quant.~Spec.~Radiat.~Transf.}}
\def\memsai{\aaref@jnl{Mem.~Soc.~Astron.~Italiana}}
\def\nphysa{\aaref@jnl{Nucl.~Phys.~A}}   
\def\physrep{\aaref@jnl{Phys.~Rep.}}   
\def\physscr{\aaref@jnl{Phys.~Scr}}   
\def\planss{\aaref@jnl{Planet.~Space~Sci.}}   
\def\procspie{\aaref@jnl{Proc.~SPIE}}   